\documentclass[%
 reprint,
 amsmath,amssymb,
 aps,
]{revtex4-2}

\usepackage{graphicx}
\usepackage{dcolumn}
\usepackage{bm}

\usepackage[caption=false]{subfig}
\usepackage{siunitx, soul, xcolor}
\usepackage{adjustbox}
\usepackage{subfig}
\usepackage{graphicx}
\usepackage[normalem]{ulem}
\usepackage{hyperref}
\hypersetup{hypertex=true,
            colorlinks=true,
            linkcolor=blue,
            anchorcolor=blue,
            citecolor=blue}   
\usepackage{gensymb}

\usepackage{xparse,xcoffins}

\ExplSyntaxOn
\NewCoffin\imagecoffin
\NewCoffin\labelcoffin

\keys_define:nn { miguel/label }
 {
  label   .tl_set:N = \l_miguel_label_tl,
  labelbox .bool_set:N = \l_miguel_label_box_bool,
  labelbox .default:n = true,
  fontsize .tl_set:N = \l_miguel_label_size_tl,
  fontsize .initial:n = \footnotesize,
  pos .choice:,
  pos/nw .code:n = \tl_set:Nn \l_miguel_label_pos_tl { left,up },
  pos/ne .code:n = \tl_set:Nn \l_miguel_label_pos_tl { right,up },
  pos/sw .code:n = \tl_set:Nn \l_miguel_label_pos_tl { left,down },
  pos/se .code:n = \tl_set:Nn \l_miguel_label_pos_tl { right,down },
  pos/n .code:n = \tl_set:Nn \l_miguel_label_pos_tl { hc,up },
  pos/w .code:n = \tl_set:Nn \l_miguel_label_pos_tl { left,vc },
  pos/s .code:n = \tl_set:Nn \l_miguel_label_pos_tl { hc,down },
  pos/e .code:n = \tl_set:Nn \l_miguel_label_pos_tl { right,vc },
  pos .initial:n = nw,
  unknown .code:n   = \clist_put_right:Nx \l_miguel_label_clist
                       { \l_keys_key_tl = \exp_not:n { #1 } }
 }
\clist_new:N \l_miguel_label_clist
\box_new:N \l_miguel_label_box
\box_new:N \l_miguel_label_image_box

\NewDocumentCommand{\xincludegraphics}{O{}m}
 {
  \group_begin:
  \tl_clear:N \l_miguel_label_tl
  \clist_clear:N \l_miguel_label_clist
  \keys_set:nn { miguel/label } { #1 }
  \tl_if_empty:NTF \l_miguel_label_tl
   {
    \miguel_includegraphics:Vn \l_miguel_label_clist { #2 }
   }
   {
    \SetHorizontalCoffin\imagecoffin
     {
      \miguel_includegraphics:Vn \l_miguel_label_clist { #2 }
     }
    \SetHorizontalCoffin\labelcoffin
     {
      \raisebox{\depth}
       {
        \bool_if:NTF \l_miguel_label_box_bool
         { \fcolorbox{white}{white}{\l_miguel_label_size_tl\l_miguel_label_tl} }
         { \l_miguel_label_size_tl\l_miguel_label_tl }
       }
     }
    \SetVerticalPole\imagecoffin{left}{-0pt+\CoffinWidth\labelcoffin/2}
    \SetVerticalPole\imagecoffin{right}{\Width-3pt-\CoffinWidth\labelcoffin/2}
    \SetHorizontalPole\imagecoffin{up}{\Height+8pt-\CoffinHeight\labelcoffin/2}
    \SetHorizontalPole\imagecoffin{down}{3pt+\CoffinHeight\labelcoffin/2}
    \use:x{\JoinCoffins\imagecoffin[\l_miguel_label_pos_tl]\labelcoffin[vc,hc]} 
    \TypesetCoffin\imagecoffin
   }
   \group_end:
 }
\NewDocumentCommand{\setlabel}{m}
 {
  \keys_set:nn { miguel/label } { #1 }
 }

\cs_new_protected:Nn \miguel_includegraphics:nn
 {
  \includegraphics[#1]{#2}
 }
\cs_generate_variant:Nn \miguel_includegraphics:nn { V }

\ExplSyntaxOff

\begin{document}

\preprint{APS/123-QED}

\title{Full-film dry transfer of MBE-grown van der Waals materials \\
}

\author{Ziling Li}
 \affiliation{Department of Physics, The Ohio State University, Columbus, Ohio 43210, United States}
\author{Wenyi Zhou}
 \affiliation{Department of Physics, The Ohio State University, Columbus, Ohio 43210, United States}
\author{Matthew Swann}
 \affiliation{Department of Physics, The Ohio State University, Columbus, Ohio 43210, United States}
\author{Vika Vorona}
 \affiliation{Department of Physics, The Ohio State University, Columbus, Ohio 43210, United States}
\author{Haley Scott}
 \affiliation{Department of Physics, The Ohio State University, Columbus, Ohio 43210, United States} 
\author{Roland K. Kawakami}%
 \email{kawakami.15@osu.edu}
 \affiliation{Department of Physics, The Ohio State University, Columbus, Ohio 43210, United States}

\begin{abstract}
Molecular beam epitaxy (MBE) has been used to create high-quality, large-scale two-dimensional van der Waals (2D vdW) materials. However, due to the strong adhesion between the substrate and deposited materials, 
the peel-off and dry transfer of MBE-grown vdW films onto other substrates has been challenging. This limits the study and use of MBE films for heterogeneous integration including stacked and twisted heterostructures. In this work, we develop a polymer-assisted dry transfer method and successfully perform full-film transfer of various MBE-grown 2D vdW materials including transition metal dichalcogenides (TMD), topological insulators (TI) and 2D magnets. In particular, we transfer air-sensitive 2D magnets, characterize their magnetic properties, and compare them with as-grown materials. The results show that the transfer technique does not degrade the magnetic properties, with the Curie temperature and hysteresis loops exhibiting similar behaviors after the transfer. Our results enable further development of heterogeneous integration of 2D vdW materials based on MBE growth.

\end{abstract}

\flushbottom
\maketitle


\section{\label{sec:level1} Introduction}
Two-dimensional van der Waals (2D vdW) materials and their heterostructures have received significant attention due to their unique electronic, optical, and magnetic properties, which make them promising candidates for applications in fields such as nanoelectronics~\cite{lemme20222d}, optoelectronics~\cite{tan20202d} and spintronics~\cite{ahn20202d}. Traditional methods for creating these heterostructures, such as mechanical exfoliation and top-down stacking~\cite{purdie2018cleaning,guo2021stacking}, have limitations in scalability and reproducibility. To address these challenges, researchers have turned to bottom-up growth techniques such as chemical vapor deposition (CVD)~\cite{cai2018chemical} and molecular beam epitaxy (MBE)~\cite{maurtua2024molecular} to produce large-scale 2D materials.

Both CVD and MBE have demonstrated the ability to grow high-quality materials. MBE specifically offers a number of advantages including precise doping control~\cite{wang2020niobium}, the ability to grow complex structures~\cite{zhou2023tuning}, non-equilibrium growth of metastable phases~\cite{zhang2016epitaxial,chen2016ordered,reis2017bismuthene}, epitaxial alignment with the substrate, \textit{in situ} growth monitoring by electron diffraction, and excellent crystal quality~\cite{dimoulas2022perspectives}. However, a critical limitation of MBE lies in its 
difficulty with fabrication of stacked and twisted heterostructures, which have been achieved with exfoliated \cite{geim_van_2013,novoselov_2d_2016,kennes_moire_2021} and CVD-grown materials \cite{watson2021transfer} through polymer-assisted dry transfer. MBE-grown films often exhibit strong adhesion to their substrates \cite{mavridi2018adhesion}, making them more difficult to peel off and dry transfer than CVD-grown films. A successful release and large-area dry transfer of MBE films while maintaining high quality would open many possibilities for stacked vdW heterostructures, hybrid structures with other materials (e.g.~oxides, nitrides), and free-standing membranes.



In this work, we introduce a novel polymer-assisted dry transfer method utilizing polycaprolactone (PCL) to realize the large-area peel-off and transfer of various MBE-grown 2D vdW materials. The technique has demonstrated full-film peeling with area limited by the substrate size (up to $5\times5$ mm$^2$ so far) and has a high area yield ($>90\%$). We have successfully transferred 2D semiconductors (MoSe$_2$, WSe$_2$), topological insulators (Bi$_2$Se$_3$, Bi$_2$Te$_3$), and 2D magnets (Fe$_3$GeTe$_2$). To test the material quality, we compare the magnetic properties of Fe$_3$GeTe$_2$ before and after the transfer and find that both the Curie temperature and the coercivity show similar behaviors. This dry transfer method not only preserves the intrinsic properties of the materials but also facilitates the development of heterogeneous integration of 2D vdW materials based on MBE growth. Our findings represent a significant advancement in the field, offering a reliable and scalable approach for fabricating high-quality 2D material heterostructures.

\begin{figure*}[ht]
\subfloat{\xincludegraphics[width=0.55\textwidth,label=(a)]{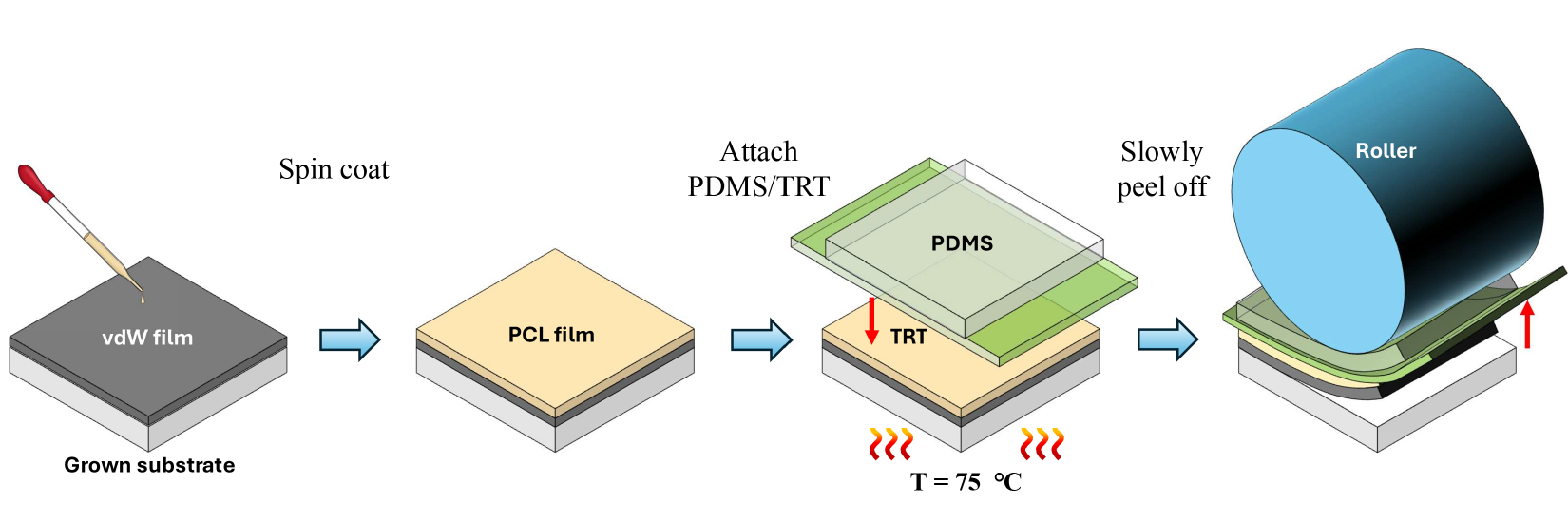}\label{fig:schematic_peel}}\hfill
\subfloat{\xincludegraphics[width=0.18\textwidth,label=(b)]
{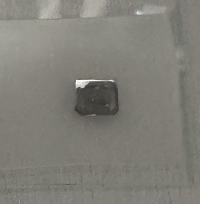}\label{fig:FGT_TRT_photo}}\hfill
\subfloat{\xincludegraphics[width=0.245\textwidth,label=(c)]
{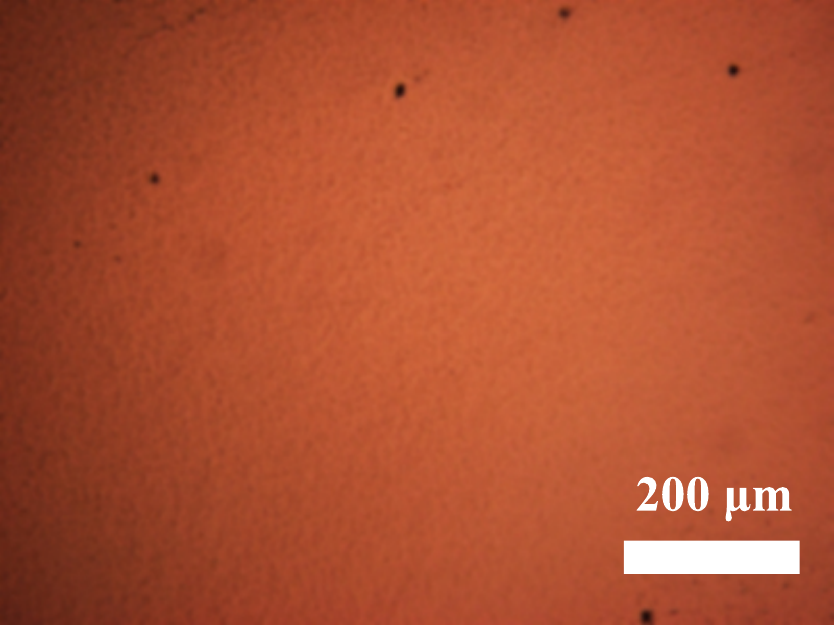}\label{fig:FGT_TRT_micro}}\hfill
\vspace{-0.01\textwidth}
\subfloat{\xincludegraphics[width=0.55\textwidth,label=(d)]{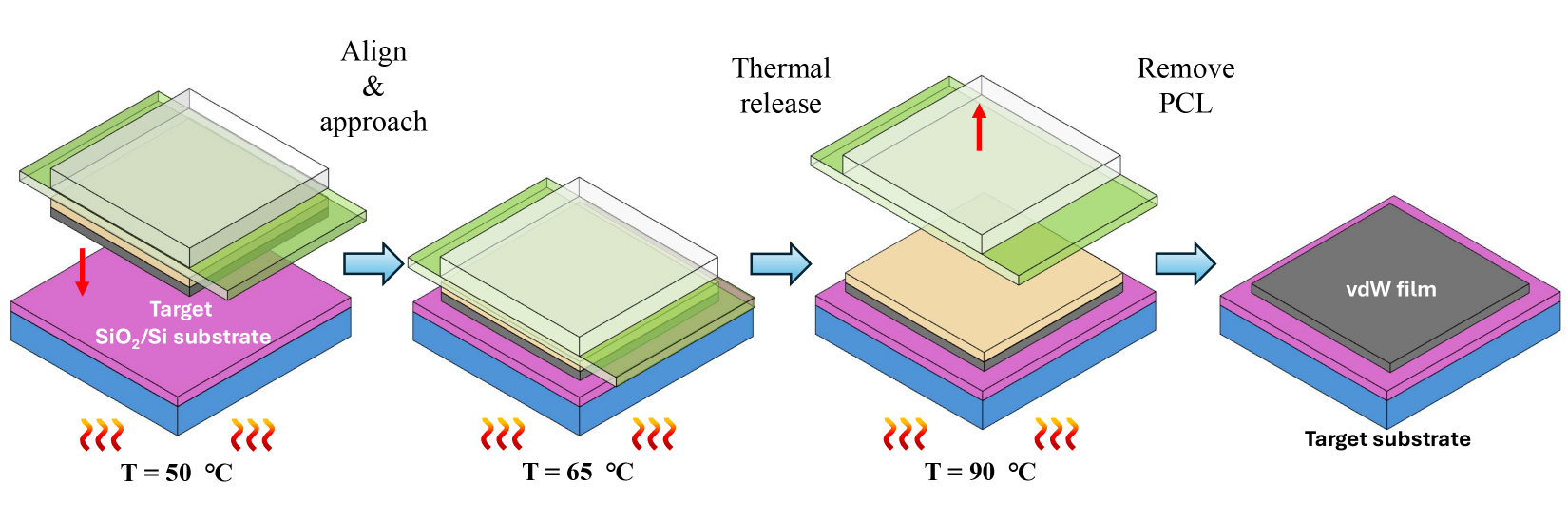}\label{fig:schematic_transfer}}\hfill
\subfloat{\xincludegraphics[width=0.183\textwidth,label=(e)]
{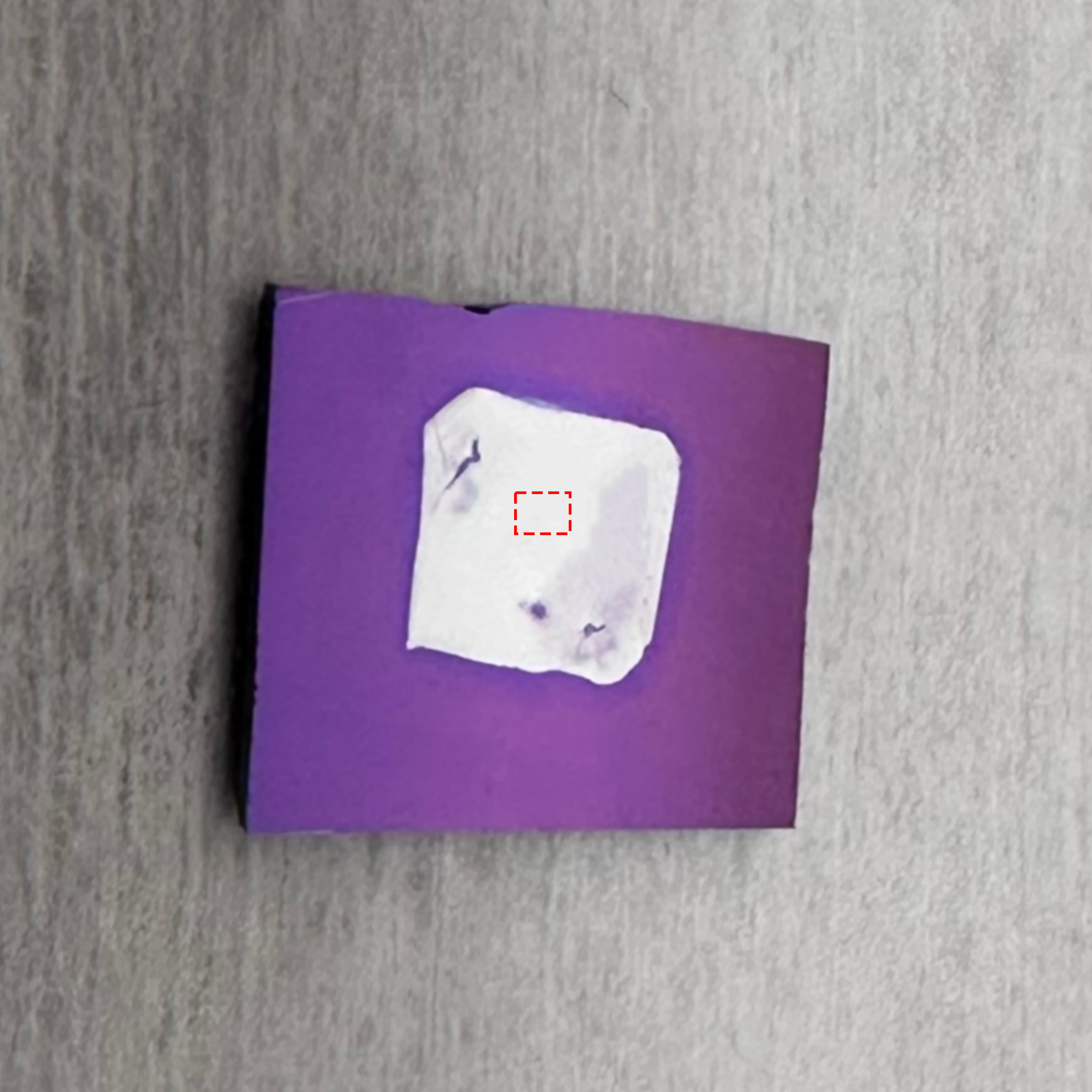}\label{fig:FGT_SiO2_photo}}\hfill
\subfloat{\xincludegraphics[width=0.245\textwidth,label=(f)]
{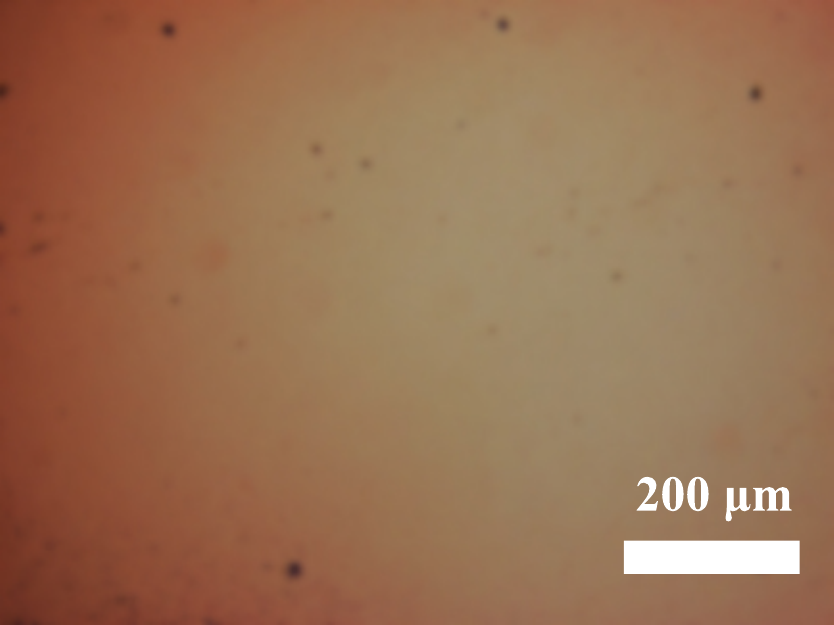}\label{fig:FGT_SiO2_micro}}\hfill
\vspace{-0.01\textwidth}
\caption{Process for PCL transfer of vdW films.
    (a) Schematic diagram of peeling off vdW film from the growth substrate.
    (b) Photograph of peeled FGT/BT film on the TRT/PDMS stamp (face up).
    (c) Microscopic optical image of the red dashed box region in (b). 
    (d) Schematic diagram of transferring vdW film onto the target substrate.
    (e) Photograph of transferred FGT/BT film on the SiO$_2$/Si substrate.
    (f) Microscopic optical image of the red dashed box region in (e).
  }
  \label{fig:MCD}
\end{figure*}

\section{\label{sec:level2}Experimental Results}
PCL (Sigma Aldrich, average M$_n$: 80000) has demonstrated strong adhesion to various 2D vdW materials and can be easily dissolved by tetrahydrofuran (THF, Sigma Aldrich, $\geq99.9\%$ anhydrous), leaving a cleaner surface compared to conventional poly(methyl methacrylate) (PMMA)~\cite{son2020strongly, park2024new}. This property makes PCL an ideal candidate for transferring vdW layers. Figure~\ref{fig:schematic_peel} illustrates the schematic of the peel-off process. A PCL/THF solution ($5\%$ by weight) is spin-coated at 1000 rpm for 1 minute and subsequently baked at 75$^\circ$C for 5 minutes to form a uniform PCL layer on top of the MBE-grown vdW film. A piece of thermal release tape (TRT, Nitto Denko Corp, Revalpha RA-95LS(N)), followed by a polydimethylsiloxane (PDMS, Gel-Pak) stamp, is gently pressed onto the PCL film to provide mechanical support. After cooling to room temperature, the entire structure is then slowly peeled off using a roller (3D printed using polylactic acid) with a small detachment angle to prevent cracking during the process. Due to the stronger adhesion between the PCL and the MBE-grown vdW film compared to the substrate, a uniform vdW film is successfully peeled, as demonstrated by the example of Fe$_3$GeTe$_2$/Bi$_2$Te$_3$ (FGT/BT) in Figures~\ref{fig:FGT_TRT_photo} and \ref{fig:FGT_TRT_micro}. 

\begin{figure}
\subfloat{\xincludegraphics[width=0.22\textwidth,label=(a)]{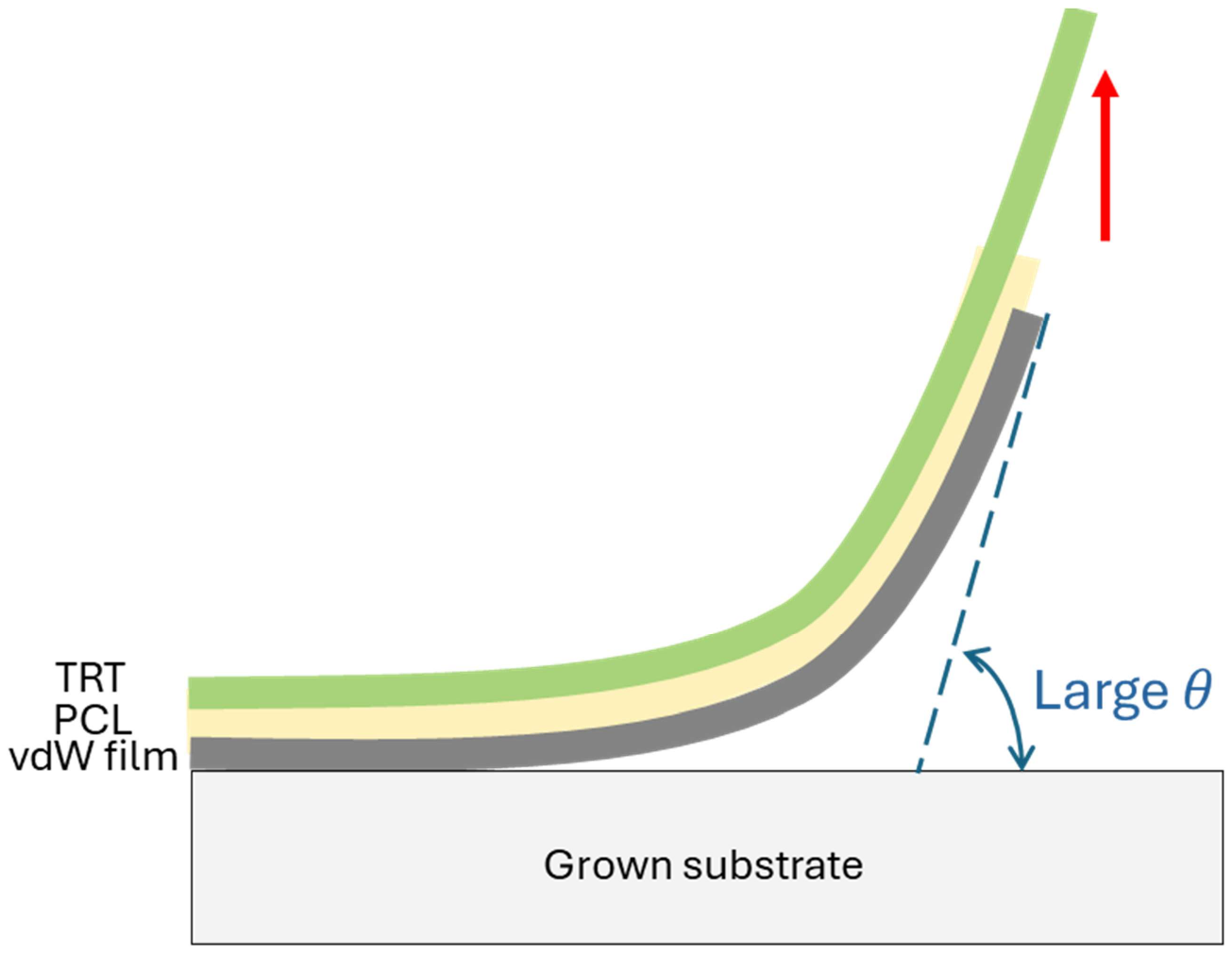}\label{fig:bad_schematic}}\hfill
\subfloat{\xincludegraphics[width=0.22\textwidth,label=(b)]
{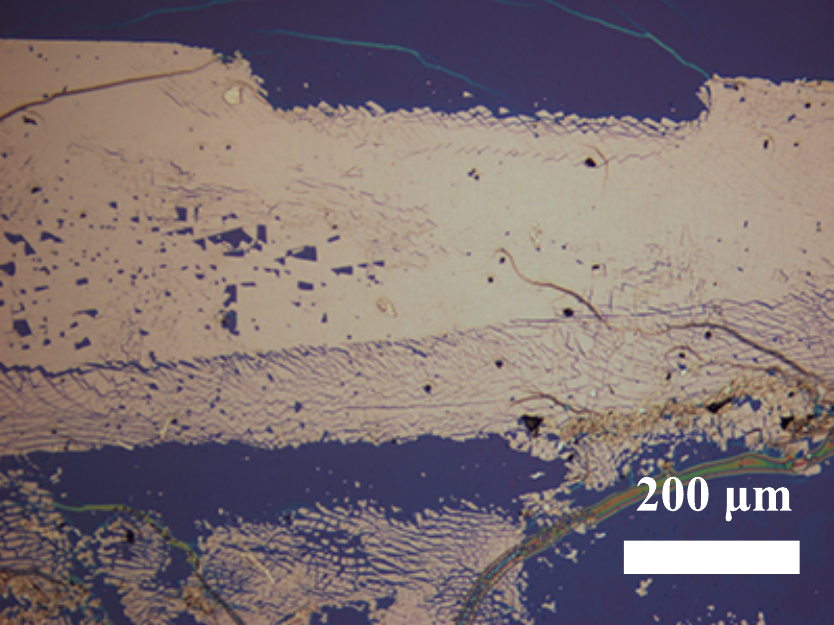}\label{fig:bad_peel}}\hfill
\subfloat{\xincludegraphics[width=0.22\textwidth,label=(c)]{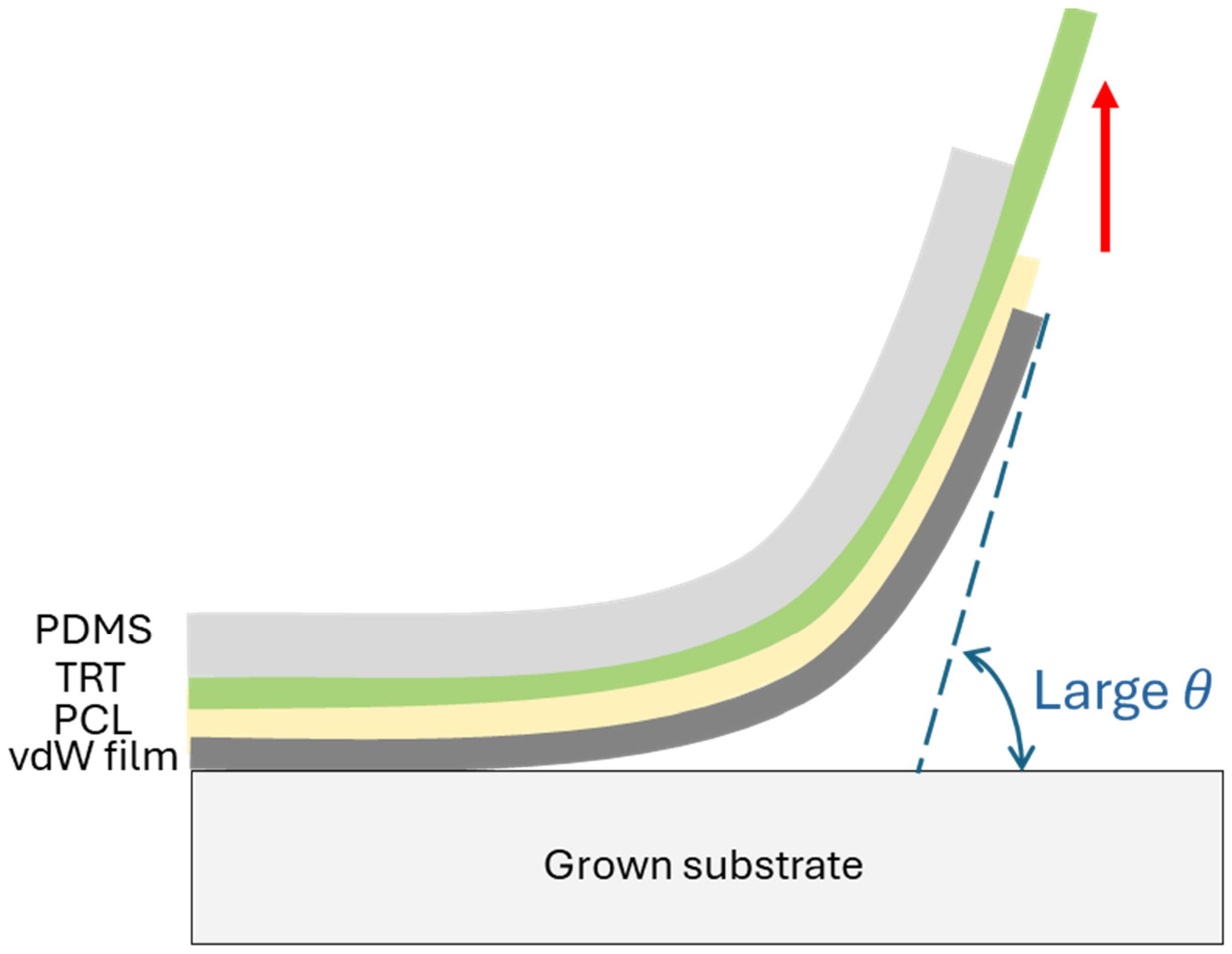}\label{fig:med_schematic}}\hfill
\subfloat{\xincludegraphics[width=0.22\textwidth,label=(d)]
{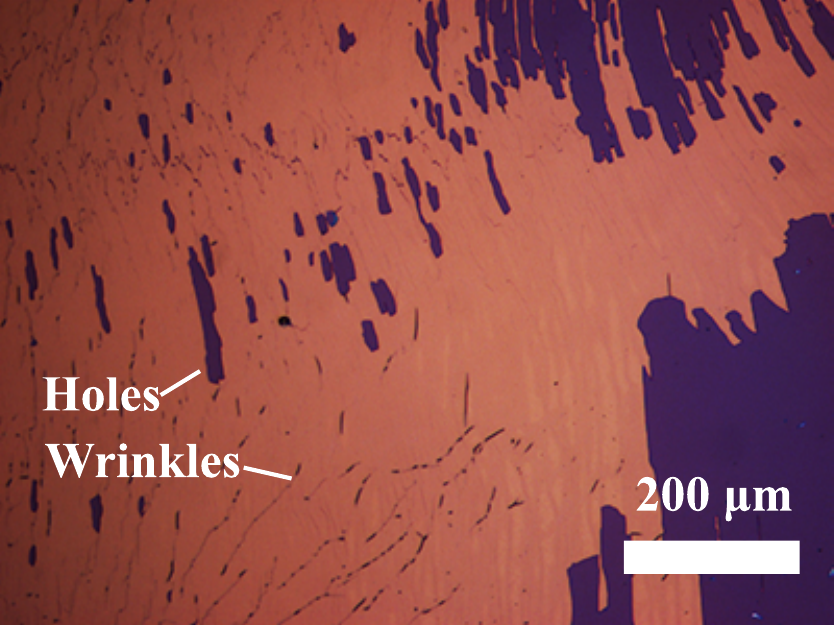}\label{fig:med_peel}}\hfill
\subfloat{\xincludegraphics[width=0.22\textwidth,label=(e)]{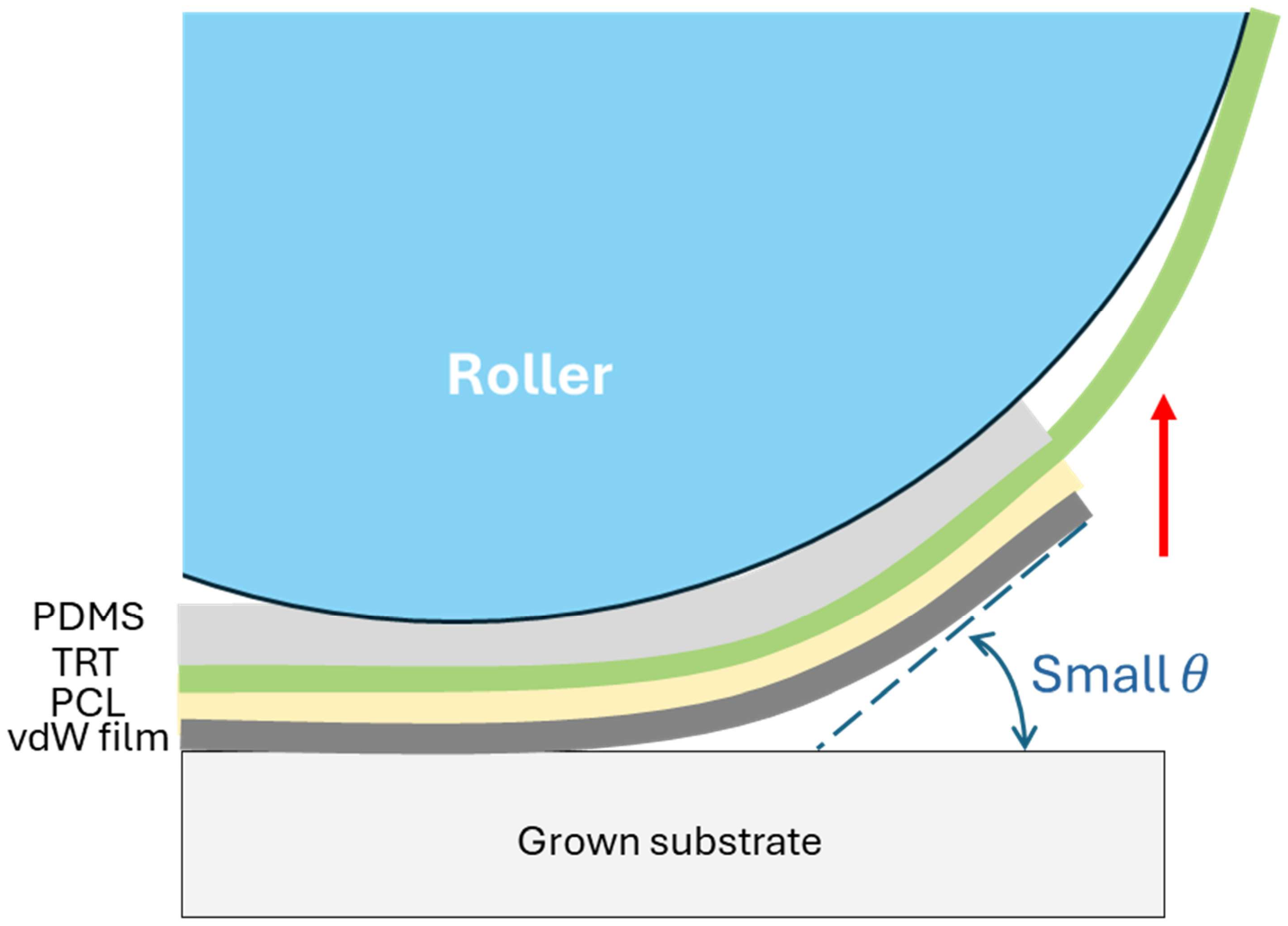}\label{fig:good_schematic}}\hfill
\subfloat{\xincludegraphics[width=0.22\textwidth,label=(f)]
{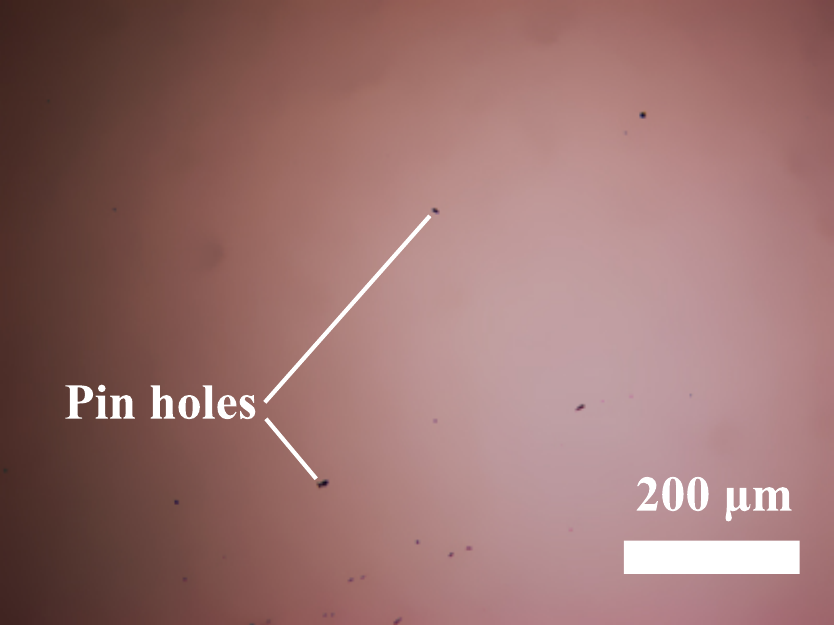}\label{fig:good_peel}}\hfill
\caption{Peel-off and transfer development.~(a) Schematic diagram using only TRT/PCL. 
    (b) Microscopic optical image of BT film transferred onto SiO$_2$/Si. 
    (c, d) Schematic diagram and microscope image by adding a PDMS support layer.
    (e, f) Schematic diagram and microscope image by adding a roller.
  }
  \label{fig:progress}
\end{figure}

To deposit the MBE film onto the target substrate, as shown in the schematic in Figure~\ref{fig:schematic_transfer}, the PDMS/TRT/PCL stamp is mounted onto a micro-manipulator for precise alignment with the target substrate. A two-step heating process is employed during the drop-down procedure. First, the stamp is approached to the target substrate at 50$^\circ$C while gradually increasing the substrate temperature to 65$^\circ$C. During this step, the PCL layer gradually melts and spreads across the contact region, minimizing the formation of bubbles between the vdW material and the substrate. This results in a flat and uniform surface of the transferred film. Once the vdW film is fully attached, the temperature is further increased to 90$^\circ$C to thermally release the PCL/vdW layer. Since 90$^\circ$C is in between the melting temperature of PCL ($\sim75^\circ$C) and TRT release temperature ($\sim105^\circ$C), only PCL will melt down and TRT residue is negligible.  Finally, the PCL layer is removed using THF solvent, leaving behind a clean and well-transferred film, as exemplified by the FGT/BT film on a SiO$_2$(thickness of 285 nm)/Si substrate in Figures~\ref{fig:FGT_SiO2_photo} and \ref{fig:FGT_SiO2_micro}.

\begin{figure*}
\centering

\subfloat{\xincludegraphics[width=0.2\textwidth,label=(a)]{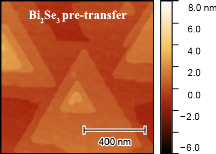}\label{fig:AFM_BS_pre}}
\subfloat{\xincludegraphics[width=0.2\textwidth,label=(b)]{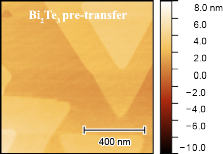}\label{fig:AFM_BT_pre}}
\subfloat{\xincludegraphics[width=0.2\textwidth,label=(c)]
{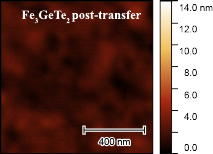}\label{fig:AFM_FGT_pre}}
\subfloat{\xincludegraphics[width=0.2\textwidth,label=(d)]
{Figure/AFM_MoSe2_pre1.pdf}\label{fig:AFM_MoSe2_pre}}
\subfloat{\xincludegraphics[width=0.2\textwidth,label=(e)]
{Figure/AFM_WSe2_pre1.pdf}\label{fig:AFM_WSe2_pre}}\hfill

\subfloat{\xincludegraphics[width=0.2\textwidth,label=(f)]{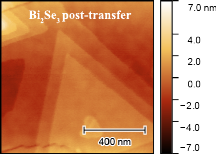}\label{fig:AFM_BS_peel}}
\subfloat{\xincludegraphics[width=0.2\textwidth,label=(g)]{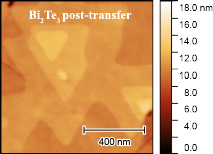}\label{fig:AFM_BT_peel}}
\subfloat{\xincludegraphics[width=0.2\textwidth,label=(h)]
{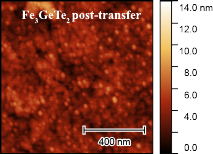}\label{fig:AFM_FGT_peel}}
\subfloat{\xincludegraphics[width=0.2\textwidth,label=(i)]
{Figure/AFM_MoSe2_peel1.pdf}\label{fig:AFM_MoSe2_peel}}
\subfloat{\xincludegraphics[width=0.2\textwidth,label=(j)]
{Figure/AFM_WSe2_peel1.pdf}\label{fig:AFM_WSe2_peel}}\hfill
\vspace{0.01\textwidth}
\subfloat{\xincludegraphics[width=0.5\textwidth,label=(k)]{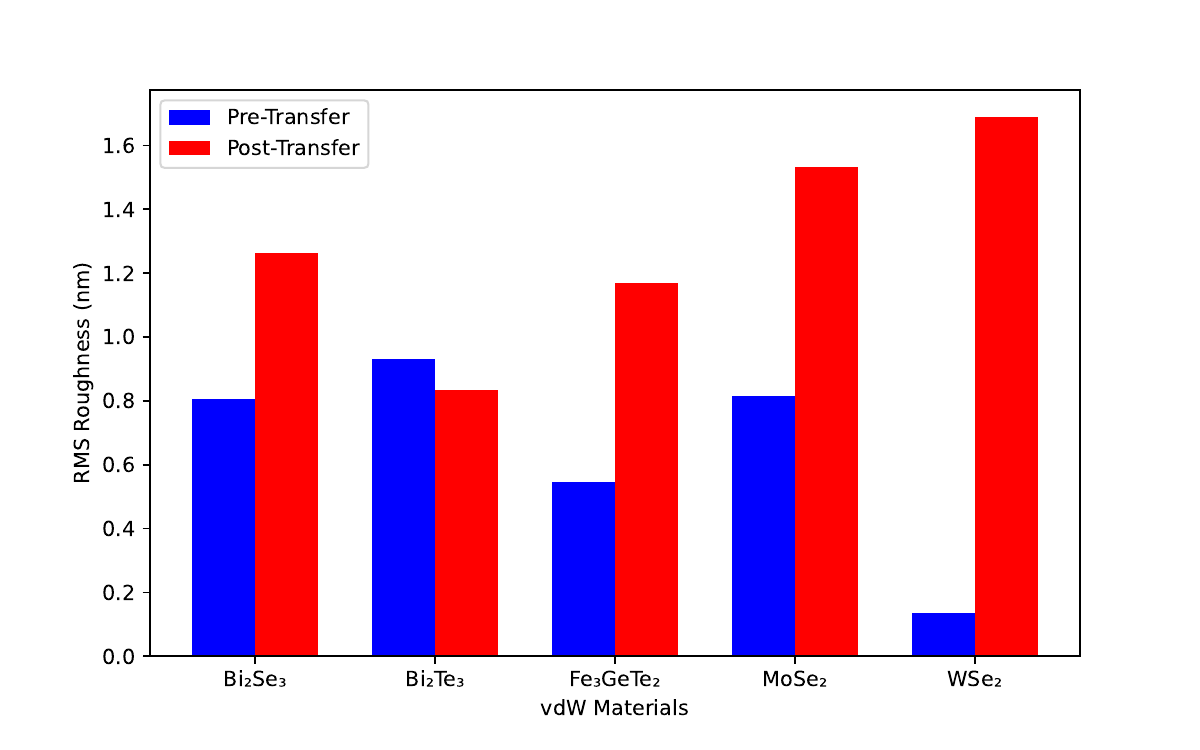}\label{fig:RMS}}\hfill
\subfloat{\xincludegraphics[width=0.45\textwidth,label=(l)]{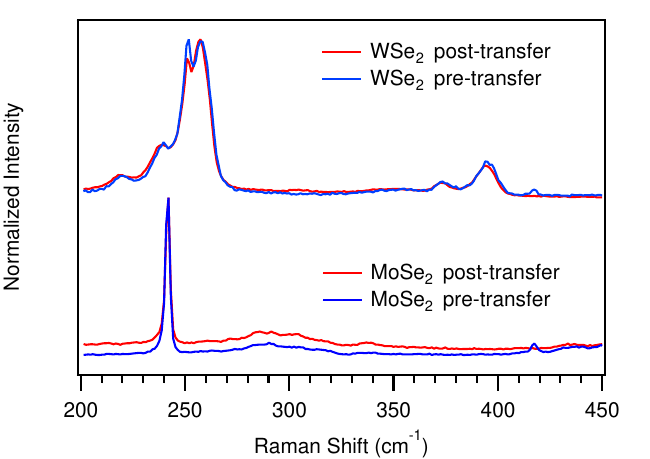}\label{fig:raman}}\hfill
\vspace{-0.01\textwidth}
\caption{Gallery of transferred vdW materials.
    (a)-(e) AFM images of as-grown Bi$_2$Se$_3$, Bi$_2$Te$_3$, Fe$_3$GeTe$_2$, MoSe$_2$ and WSe$_2$ on c-sapphire substrates. 
    (f)-(j) AFM images of transferred Bi$_2$Se$_3$, Bi$_2$Te$_3$, Fe$_3$GeTe$_2$, MoSe$_2$ and WSe$_2$ on SiO$_2$/Si substrates.
    (k) Statistics of RMS surface roughness of vdW films before (blue) and after (red) PCL transfer.
    (l) Raman spectra of MoSe$_2$ and WSe$_2$ before (blue) and after (red) PCL transfer. The intensities are normalized for better comparison.
  }
  \label{fig:gallery}
\end{figure*}

For developing this process, we have systematically tested various configurations of the lifting components. When only PCL and TRT are utilized, as illustrated in Figure~\ref{fig:bad_schematic}, the peeled-off and transferred BT film exhibits patchy coverage (Figure~\ref{fig:bad_peel}). 
Introducing a PDMS layer atop the TRT/PCL stack as a supporting layer (Figure~\ref{fig:med_schematic}) significantly improves the peel-off and transfer processes, yielding BT films with enhanced coverage (Figure~\ref{fig:med_peel}). However, residual wrinkles, cracks, and holes persist in the transferred film. 
Attributing this improvement to the added rigidity provided by the PDMS during peel-off, we further incorporate a rigid 3D-printed cylindrical roller (Figure~\ref{fig:good_schematic}) with a large radius of curvature ($\sim20$ cm). A small detachment angle $\theta$ is maintained to minimize tensile and buckling strain during peeling, ultimately achieving a uniform and crack-free transfer, as demonstrated in Figure~\ref{fig:good_peel}. 
In practice, our roller is not a full cylinder as depicted schematically in Figure~\ref{fig:schematic_peel}, but is a smaller segment of the cylindrical surface (a few cm in length) to accommodate a large radius curvature for a small detachment angle.

To explore the universality of PCL transfer, we apply the PCL transfer to different MBE-grown vdW materials including TMD MoSe$_2$ and WSe$_2$, TI Bi$_2$Se$_3$ and Bi$_2$Te$_3$, and 2D magnet Fe$_3$GeTe$_2$ (See Supplementary Material for details on the MBE growth \cite{SupplMat}). All the materials show good transfer results which indicates the PCL transfer method can be employed for a variety of 2D vdW materials. To further evaluate the surface quality of the transferred film, we performed atomic force microscopy (AFM) measurements on uncapped vdW films before (Figures~\ref{fig:AFM_BS_pre}-\ref{fig:AFM_WSe2_pre}) and after transfer (Figures~\ref{fig:AFM_BS_peel}-\ref{fig:AFM_WSe2_peel}). Figure~\ref{fig:RMS} shows that the samples have comparable root-mean-squared (RMS) roughness before and after transfer, ranging from 0.8 nm to 1.6 nm. The increase of RMS is about 1 nm, which is better than traditional PMMA-based transfer ($\sim 5$ nm) and is comparable to optimized transfer of CVD-grown materials \cite{mondal2024low, han2014clean}.

To examine the material quality of the transferred film, we also measure Raman spectra of MoSe$_2$ and WSe$_2$ before and after PCL transfer using a Renishaw inVia confocal Raman microscope. Raman directly probes the lattice vibrational modes for a given material and is sensitive to the crystallinity of the film. It is expected that as the crystalline quality deteriorates, the linewidth will increase and the peak positions may change~\cite{lee_spatially_2017,mignuzzi_effect_2015,neumann_raman_2015}. Spectra are collected using a 633 nm excitation source transmitted through a 50$\times$ lens located above the sample, along with a 1800 lines/mm grating and a CCD detector. Raman spectroscopy of MoSe$_2$ and WSe$_2$ before and after transfer yields nearly identical spectra. The widths and peak positions before and after remain largely unchanged, with only a small decrease in intensity of the e$_g$ peak for WSe$_2$ at 251 cm$^{-1}$ relative to other modes \cite{del_corro_excited_2014}, confirming the films' crystalline quality remains stable through the transfer process. 

\begin{figure}
\subfloat{\xincludegraphics[width=0.24\textwidth,label=(a)]{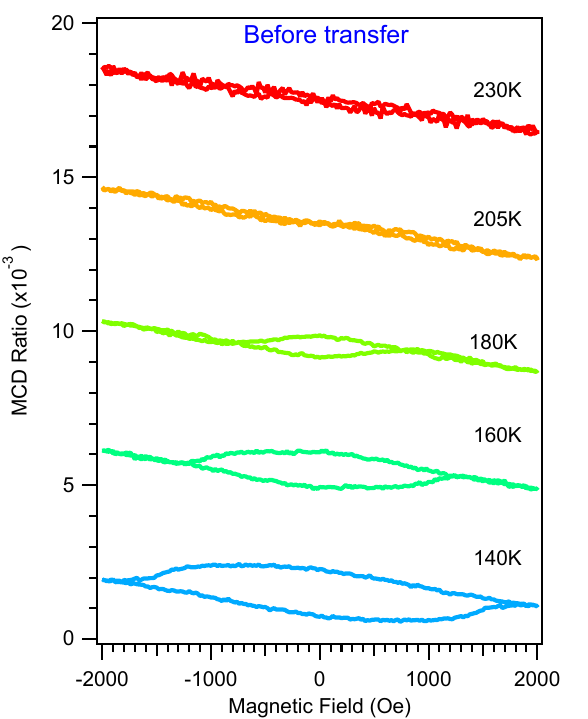}\label{fig:MCD_bef}}\hfill
\subfloat{\xincludegraphics[width=0.24\textwidth,label=(b)]{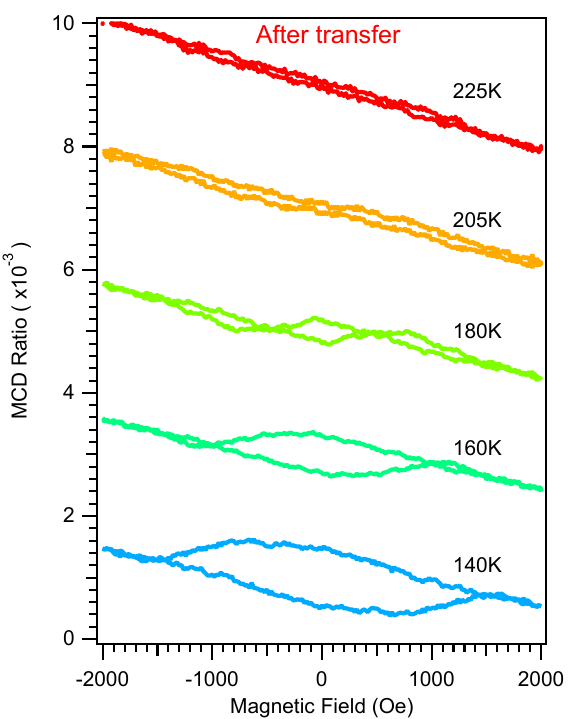}\label{fig:MCD_aft}}\hfill
\vspace{-0.01\textwidth}
\subfloat{\xincludegraphics[width=0.24\textwidth,label=(c)]{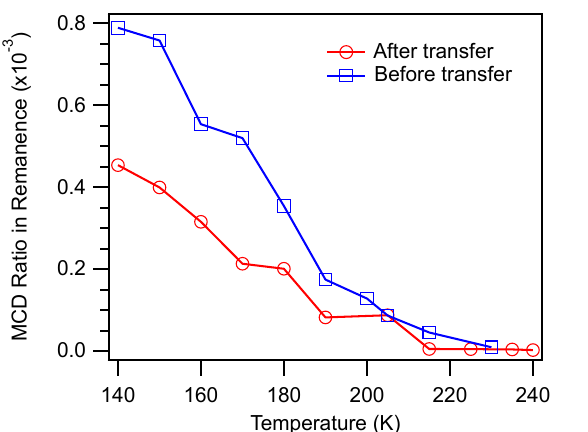}\label{fig:MCD_R-T}}\hfill
\subfloat{\xincludegraphics[width=0.24\textwidth,label=(d)]{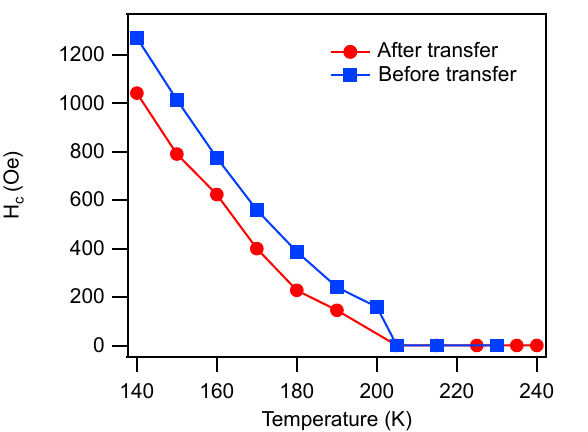}\label{fig:MCD_Hc-T}}\hfill
\vspace{-0.01\textwidth}
\caption{Magnetic properties of FGT/BT films: (a) Temperature dependence of MCD loops for FGT/BT sample before transfer. (b) Temperature dependence of MCD loops for FGT/BT sample after transfer.(c) Temperature dependence of extracted MCD ratio in remanence. (d) Temperature dependence of extracted coercive field $H_c$.
  }
  \label{fig:MCD}
\end{figure}

Finally, we demonstrate the transfer of a complex 2D magnet/TI heterostructure \cite{zhou2023tuning} and compare the magnetic properties before and after transfer.
The FGT/BT film is capped with 10 nm of Te to prevent oxidation after MBE growth (See supplementary\cite{SupplMat}). We transfer the Te/FGT/BT film from the c-sapphire substrate to SiO$_2$/Si substrate using the PCL transfer method in air. To investigate the influence of transfer on the magnetic properties, we perform reflective magnetic circular dichroism (MCD) to measure out-of-plane magnetic hysteresis loops of the Te/FGT/BT heterostructures before and after transferring. Samples are loaded into an optical cryostat (Advanced Research Systems, DMX-20-OM) to control the temperature. We utilize a continuous wave 532 nm laser with power of 100 $\mu$W focused onto a $\sim$3\,$\mu$m-diameter spot using a $50\times$ microscope objective at normal incidence. The polarization of the incident light is modulated between left-circularly polarized (LCP) and right-circularly polarized (RCP) by a photoelastic modulator (Hinds Instrument) at a frequency of 59 kHz. The reflected beam intensity ($I$) is measured by a photodiode detector (Thorlabs PDA36A2) and the MCD ratio ${(I_{RCP}-I_{LCP})}/{(I_{RCP}+I_{LCP})}$ is extracted by a lock-in amplifier (Signal Recovery 7270).  A magnetic field is applied by a vector electromagnet to apply an out-of-plane magnetic field. The magnet is controlled by a data acquisition card (National Instruments PCIe-6323), which enables a relatively high ramp rate of 400 Oe/s to avoid drift of the MCD signal. The presented MCD hysteresis loops are typically averaged over 10-20 scans, and the loops have been antisymmetrized, i.e.~$MCD_{up}^{anti}(H)=\frac{1}{2}[MCD_{up}(H)-MCD_{down}(-H)]$, $MCD_{down}^{anti}(H)=\frac{1}{2}[MCD_{down}(H)-MCD_{up}(-H)]$, where ``up'' and ``down'' refer to the field sweep direction.

Figure~\ref{fig:MCD} compares the magnetic properties before and after the transfer. Prior to the transfer, out-of-plane magnetic hysteresis loops are measured by MCD at a series of temperatures (Figure~\ref{fig:MCD_bef}). At 140 K, the hysteresis loop exhibits a large coercivity, $H_c$, of $\sim1200$ Oe. At lower temperatures, the sample cannot be fully saturated with the electromagnet ($\sim2000$ Oe maximum field). Above 140 K, $H_c$ decreases with increasing temperature, and the magnetic hysteresis loop disappears by $\sim200$ K. Figure \ref{fig:MCD_aft} shows the corresponding data after the FGT/BT has been transferred onto a SiO$_2$/Si substrate. Here, the behavior is similar to the as-grown sample, with the magnetization preserving its perpendicular orientation and the magnetic hysteresis loop disappearing by $\sim200$ K. However, $H_c$ is slightly reduced compared to the as-grown sample.

Figures~\ref{fig:MCD_R-T} and \ref{fig:MCD_Hc-T} plot the MCD ratio in remanence and coercivity, respectively, as a detailed function of temperature and provide a comparison before and after the transfer. The most noticeable difference is the magnitude of the MCD ratio in remanence, which is likely a consequence of the optical properties as opposed to changes in the magnetic properties. For instance, the presence of the $\sim285$ nm thick SiO$_2$ layer on the Si substrate creates an optical interference effect that could modify the magneto-optic response. Nevertheless, in both cases the MCD signal disappears above $\sim200$ K, signifying that the Curie temperature is largely preserved. For the coercivity before and after the transfer, $H_c$ decreases with increasing temperature and disappears at the Curie temperature. However, the magnitudes of $H_c$ are slightly smaller after the transfer. In summary, the Curie temperature and perpendicular magnetization orientation remain unchanged after transferring, and although there are small changes to $H_c$ due to the transfer, we are highly encouraged that the magnetic properties are largely preserved overall.

\section{\label{sec:level4}Conclusion}
In conclusion, we have developed a polymer-assisted dry transfer method utilizing PCL to enable the robust, large-area transfer of MBE-grown van der Waals materials, overcoming the critical challenge of strong substrate adhesion inherent to epitaxial films. By optimizing the interplay between PCL adhesion, thermal release tape, and PDMS mechanical support, we have achieved a high area yield transfer of air-sensitive 2D magnets, topological insulators, and semiconductors onto arbitrary substrates. Atomic force microscopy and Raman spectroscopy confirm the preservation of the structural integrity, with the increase in RMS roughness from the transfer limited to $\sim1$ nm, surpassing conventional PMMA-based methods. Crucially, magnetic circular dichroism measurements reveal negligible changes in the Curie temperature of FGT/BT heterostructures, underscoring the method’s compatibility with delicate magnetic and electronic states. The universality of this technique, validated across diverse material systems, bridges the gap between high-quality MBE growth and functional device integration, opening opportunities for twisted heterostructures, hybrid quantum devices, and freestanding membranes. Future efforts could focus on scaling the process to wafer-sized substrates and exploring novel heterostructures for spintronic, quantum, and optoelectronic applications. Our results establish a robust platform for advancing the heterogeneous integration of 2D materials, paving the way for next-generation van der Waals technologies.

\section{Data availability statement}

The data that support the findings of this study are available upon reasonable request from the authors.

\section{Conflicts of interest}

The authors declare no conflict of interest.

\begin{acknowledgments}
The research was primarily supported by the U.S. Department of Energy, Office of Science, Basic Energy Sciences under award number~DE-SC0016379 (Z.L.~and R.K.K.). W.Z.~acknowledges support from the Center for Energy Efficient Magnonics, an Energy Frontiers Research Center funded by the U.S.~Department of Energy, Office of Science, Basic Energy Sciences under award number DE-AC02-76SF00515 and the AFOSR/MURI project 2DMagic under award number FA9550-19-1-0390 (for MBE growth, MCD, XRD, and AFM of TIs and 2D magnets). M.T.S.~acknowledges support from Intel CAF\'E (for MBE growth, Raman, and AFM of TMDs).
Partial funding for shared facilities used in this research was provided by the Center for Emergent Materials: an NSF MRSEC under award number DMR-2011876.

\end{acknowledgments}


\bibliography{apssamp}

\end{document}